\documentclass[twocolumn,preprintnumbers,prd,nofootinbib,superscriptaddress]{revtex4-1}

\usepackage{amsmath,amssymb,amsthm,stackrel}
\usepackage{graphicx}
\usepackage{changepage}
\usepackage{hyperref,url,breakurl}
\usepackage{mathtools,xparse}
\usepackage{verbatim}
\usepackage{dsfont}
\usepackage{xcolor}

\theoremstyle{definition}

\newtheorem*{definition*}{Definition}

\def\ket#1{| #1 \rangle}

\def\app#1#2{%
  \mathrel{%
    \setbox0=\hbox{$#1\sim$}%
    \setbox2=\hbox{%
      \rlap{\hbox{$#1\propto$}}%
      \lower1.1\ht0\box0%
    }%
    \raise0.25\ht2\box2%
  }%
}

\begin{document}

\title{The Multiverse in an Inverted Island}

\author{Kevin Langhoff}
\affiliation{Berkeley Center for Theoretical Physics, Department of Physics, University of California, Berkeley, CA 94720, USA}
\author{Chitraang Murdia}
\affiliation{Berkeley Center for Theoretical Physics, Department of Physics, University of California, Berkeley, CA 94720, USA}
\author{Yasunori Nomura}
\affiliation{Berkeley Center for Theoretical Physics, Department of Physics, University of California, Berkeley, CA 94720, USA}
\affiliation{Theoretical Physics Group, Lawrence Berkeley National Laboratory, Berkeley, CA 94720, USA}
\affiliation{Kavli Institute for the Physics and Mathematics of the Universe (WPI), The University of Tokyo Institutes for Advanced Study, Kashiwa 277-8583, Japan}

%-------------------------------------------------------------------------

\begin{abstract}
We study the redundancies in the global spacetime description of the eternally inflating multiverse using the quantum extremal surface prescription.
We argue that a sufficiently large spatial region in a bubble universe has an entanglement island surrounding it.
Consequently, the semiclassical physics of the multiverse, which is all we need to make cosmological predictions, can be fully described by the fundamental degrees of freedom associated with certain finite spatial regions.
The island arises due to mandatory collisions with collapsing bubbles, whose big crunch singularities indicate redundancies of the global spacetime description.
The emergence of the island and the resulting reduction of independent degrees of freedom provides a regularization of infinities which caused the cosmological measure problem.
\end{abstract}

%-------------------------------------------------------------------------

\maketitle

\makeatletter
\def\l@subsection#1#2{}
\def\l@subsubsection#1#2{}
\makeatother
%\tableofcontents

%-----------------------------------------------------------------------

\section{Introduction}
\label{sec:intro}

In the last two decades or so, we have learned a lot about the origin of spacetime in quantum gravity.
A key concept is holography~\cite{tHooft:1993dmi,Susskind:1994vu,Maldacena:1997re,Bousso:2002ju}, which states that a fundamental description of quantum gravity resides in a spacetime, often non-gravitational, whose dimension is lower than that of the bulk spacetime.
This concept has been successfully applied to understanding the dynamics of an evaporating black hole, in particular to address the information problem~\cite{Hawking:1976ra}; for recent reviews, see Refs.~\cite{Almheiri:2020cfm,Nomura:2020ewg,Raju:2020smc}.

There are two distinct approaches to implementing the idea of holography.
One is to start from global spacetime of general relativity and identify independent quantum degrees of freedom~\cite{Penington:2019npb,Almheiri:2019psf,Almheiri:2019hni} using the quantum extremal surface (QES) prescription~\cite{Ryu:2006bv,Hubeny:2007xt,Faulkner:2013ana,Engelhardt:2014gca}.
When applying this prescription to a black hole, the existence of the interior is evident, whereas understanding unitary evolution requires non-perturbative gravitational effects~\cite{Penington:2019kki,Almheiri:2019qdq}.
The other approach is to begin with a description that is manifestly unitary (if all the relevant physics is included in the infrared) and understand how the picture of global spacetime emerges~\cite{Papadodimas:2012aq,Papadodimas:2013jku,Papadodimas:2015jra,Nomura:2018kia,Nomura:2019qps,Nomura:2020ska}.
Specifically, in this approach the interior of an evaporating black hole arises as a collective phenomenon of soft (and radiation) modes~\cite{Nomura:2018kia,Nomura:2019qps,Nomura:2020ska,Maldacena:2013xja}.
While the two approaches appear radically different at first sight, they are consistent with each other in the common regime of applicability~\cite{Langhoff:2020jqa,Neuenfeld:2021bsb}.

In this paper, we study the eternally inflating multiverse using the first approach which begins with global spacetime.
A key assumption is that for a partial Cauchy surface $R$ in a weakly gravitating region, we can use the QES prescription~\cite{Engelhardt:2014gca}.
In particular, the von~Neumann entropy of the microscopic degrees of freedom associated with the region $R$ is given by the island formula~\cite{Almheiri:2019hni}
\begin{equation}
  S({\bf R}) = {\rm min} \stackbin[I]{}{\rm ext} S_{\rm gen}(I \cup R),
\label{eq:S_R}
\end{equation}
where $I$ is a partial Cauchy surface spacelike separated from $R$.%
\footnote{In this paper, $I$ refers to a spacelike codimension-1 surface.
Although it is more standard to refer to the domain of dependence of $I$, $D(I)$, as the island, we also refer to $I$ as an island in this paper.}
Here, the boldface symbol ${\bf R}$ in the left-hand side is to emphasize that $S({\bf R})$ is the microscopic von~Neumann entropy of the fundamental degrees of freedom, while
\begin{equation}
  S_{\rm gen}(X) = \frac{{\cal A}(\partial X)}{4G_{\rm N}} + S_{\rm bulk}(X)
\label{eq:S_gen}
\end{equation}
is the generalized entropy for partial Cauchy surface $X$ calculated in bulk semiclassical theory, where ${\cal A}(\partial X)$ is the area of the boundary $\partial X$ of $X$, and $S_{\rm bulk}(X)$ is the von~Neumann entropy of the reduced density matrix of $X$ calculated in the semiclassical theory.

In this work, we show that when $R$ is a sufficiently large region on a late time hypersurface in a bubble universe, an island $I$ appears which encloses the bubble universe.
Given that the semiclassical physics in $I$ is fully reconstructed using the fundamental degrees of freedom in $R$, this implies that the full semiclassical physics of the multiverse needed to make cosmological predictions is encoded in the fundamental degrees of freedom of the region $R$, which has a finite volume!

While one might feel that this is too drastic a conclusion, in some respects it is not.
Even for a black hole, the interior region described as an island $I$ can have an ever increasing spatial volume, which can even be infinite for an eternal black hole~\cite{Christodoulou:2014yia,Christodoulou:2016tuu}.
However, in quantum gravity, the number of independent states associated with this region is bounded by the exponential of the entropy of the system.
This is because exponentially small overlaps between semiclassically orthogonal states lead to a drastic reduction in the number of basis states~\cite{Langhoff:2020jqa,Jafferis:2017tiu,Marolf:2020xie,Chakravarty:2020wdm}.
What happens in the multiverse is an ``inside-out'' version of the black hole case.
As anticipated in Refs.~\cite{Nomura:2011dt,Bousso:2011up,Nomura:2011rb}, this allows us to address the cosmological measure problem~\cite{Guth:2000ka,Vilenkin:2006xv,Linde:2014nna,Freivogel:2011eg,Hebecker:2020egx} associated with the existence of an infinitely large spacetime at the semiclassical level.

\subsubsection*{Entanglement Castle}

In the black hole case, the region $R$ encloses $I$, so $I$ looks geographically like an island. 
However, in our setup, $I$ encloses $R$ so it no longer appears as an island.
Thus, we call $I$ an {\it inverted island}.

\begin{figure}[t]
\centering
\vspace{-0.6cm}
  \includegraphics[clip,width=0.9\columnwidth]{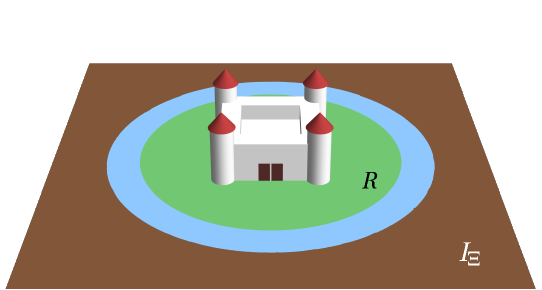}
\vspace{0.1cm}
\caption{The multiverse as an entanglement castle.
On a given Cauchy surface $\Xi$, the physics of the multiverse can be described by the fundamental degrees of freedom associated with the region $R \cup (\overline{R \cup I_\Xi})$, where $I_\Xi = D(I) \cap \Xi$ with $I$ being the (inverted) island of a partial Cauchy surface $R$.}
\label{fig:castle}
\end{figure}

The geography for a Cauchy surface $\Xi$ containing $R$ is depicted in Fig.~\ref{fig:castle}.
It is customary to treat the regions $R$ and $I$ as ``land'' and everything else as ``water.''
Following this convention, $\Xi$ has a central land $R$ surrounded by a moat $\overline{R \cup I_\Xi}$ which separates $R$ from $I_\Xi$, where $I_\Xi = D(I) \cap \Xi$.
To describe the multiverse at the semiclassical level, one only needs fundamental degrees of freedom associated with the complement of $I_\Xi$ on $\Xi$,  $\overline{I_\Xi} = R \cup (\overline{R \cup I_\Xi})$.
This is the region corresponding to the castle---the multiverse lives in an {\it entanglement castle}.

\subsubsection*{Relation to Other Work}

Entanglement islands in cosmological spacetimes have been discussed in the context of toy models, e.g., models in which a nongravitational bath is entangled with a gravitational system as well as models in lower dimensional gravity~\cite{Dong:2020uxp,Krishnan:2020fer,Chen:2020tes,Hartman:2020khs,Hartman:2020khs,Murdia:2020iac,Balasubramanian:2020xqf,Sybesma:2020fxg,Manu:2020tty,Choudhury:2020hil,Geng:2021wcq,Aalsma:2021bit}.
In this paper, we study them in a realistic scenario of eternal inflation.

Several holographic descriptions of the multiverse have been proposed~\cite{Freivogel:2006xu,Susskind:2007pv,Sekino:2009kv,Nomura:2011dt,Bousso:2011up,Nomura:2011rb,Vilenkin:2013loa,Hartle:2016tpo,Nomura:2016ikr}, mostly to address the measure problem.
These correspond to the unitary description of a black hole, although the issue of unitarity at the fundamental level is not quite clear in cosmology.

\subsubsection*{Outline of the Paper}

In Section~\ref{sec:review}, we review the eternally inflating multiverse and describe some basic assumptions employed in our analysis.
In Section~\ref{sec:S_bulk}, we discuss how bulk entanglement necessary for the emergence of an island can arise from accelerating domain walls, which are pervasive in the eternally inflating multiverse.

Section~\ref{sec:main} is the main technical part of the paper, in which we show that a sufficiently large region $R$ in a bubble universe has an inverted island that surrounds $R$.
Implications of this result for the multiverse are discussed in Section~\ref{sec:evol}.
Finally, Section~\ref{sec:concl} is devoted to conclusions.

\section{The Eternally Inflating Multiverse in Global Spacetime}
\label{sec:review}

In this paper, we are concerned with eternally inflating cosmology.
Eternal inflation occurs when the theory possesses a metastable vacuum which has a positive vacuum energy and small decay rates to other vacua~\cite{Gott:1982zf,Guth:1982pn}.
If the universe sits in such a vacuum at some moment, there will always be some spacetime region that remains inflating for an arbitrarily long time.

This scenario of eternal inflation is naturally realized in the string landscape~\cite{Bousso:2000xa,Kachru:2003aw,Susskind:2003kw,Douglas:2003um}.
In the string landscape, the number of local minima of the potential, i.e.\ false vacua, is enormous.
Vacuum energies at these minima can be either positive or negative.
Since exactly vanishing vacuum energy requires an infinite amount of fine-tuning, we expect that it is realized only in supersymmetric vacua.

Spacetime regions in different vacua are created by nucleation of bubbles, each of which can be viewed as a separate universe.
We assume that bubble nucleation occurs through Coleman-De~Luccia tunneling~\cite{Coleman:1980aw}, although we expect that our results also apply to other vacuum transition mechanisms such as the thermal Hawking-Moss process~\cite{Hawking:1981fz,Weinberg:2006pc}.

As explained in the introduction, we begin with the global spacetime picture, which is the infinitely large multiverse with a fractal structure generated by continually produced bubbles.
We assume that the global quantum state on a Cauchy surface is pure.
We are interested in studying the existence and location of the island corresponding to a partial Cauchy surface $R$ in the global multiverse. 

To address this problem, we focus on a particular bubble, which we call the {\it central bubble}.
We assume that the central bubble is formed in a parent de~Sitter (dS) bubble.
After being nucleated, it undergoes collisions with other bubbles~\cite{Guth:1982pn}.
Let us follow a timelike geodesic to the future along (and outside) the bubble wall separating the central bubble from other bubbles.
The last bubble that this geodesic encounters must be either an anti-de~Sitter (AdS) bubble or a supersymmetric Minkowski bubble, or else the geodesic still has an infinite amount of time to encounter another bubble.

\begin{figure}[t]
\centering
\vspace{0.1cm}
  \includegraphics[clip,width=0.9\columnwidth]{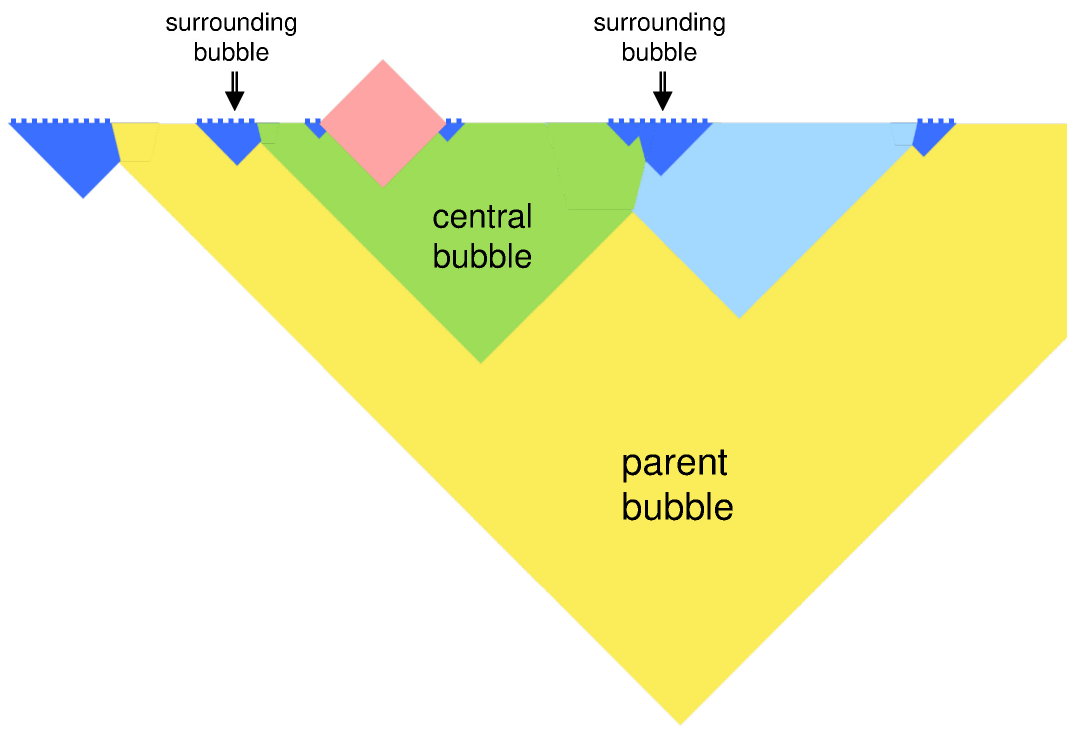}
\caption{A sketch of the Penrose diagram of the multiverse.
We focus on an arbitrarily chosen bubble, which we call the central bubble.
The central bubble is nucleated in a parent dS bubble and is surrounded by collapsing AdS bubbles which collide with it at late times.}
\label{fig:multiverse}
\end{figure}

We assume that the last bubbles such geodesics encounter are all AdS bubbles and call them {\it surrounding AdS bubbles}.
Since AdS bubbles generally end up with big crunch singularities~\cite{Coleman:1980aw}, they are collapsing bubbles.
Note that the choice of the central bubble was arbitrary, so all the bubbles have the feature of being surrounded by collapsing AdS bubbles.
A typical example of the spacetime structure described here is illustrated in Fig.~\ref{fig:multiverse}.
(We have omitted an infinite number of bubbles that form a fractal structure in the asymptotic future infinity which are not relevant for the discussion here.)

We postulate that the cosmological history we study takes place in the semiclassical regime.
This implies that the characteristic energy scale $E$ of the potential is sufficiently smaller than the cutoff scale, and hence the Planck scale.
On the other hand, in the string landscape we expect that this energy scale is not much smaller than the string scale, e.g., $E \sim O(10^{-5}~\mbox{--}~10^{-1})/l_{\rm P}$, where $l_{\rm P}$ is the Planck length.
Note, however, that some of these bubbles could be associated with much smaller energy scales by selection effects.
For instance, the bubble universe that we live in has a vacuum energy much smaller than the naive value of $O(E^4)$~\cite{Weinberg:1987dv,Banks:1984cw,Linde:1984ir}.

\section{Bulk Entanglement from Accelerating Domain Walls}
\label{sec:S_bulk}

In this section, we discuss the possible origin of bulk entanglement $S_{\rm bulk}$ leading to an island in eternally inflating spacetime.
As discussed in Ref.~\cite{Hartman:2020khs}, an island cannot be created by $S_{\rm bulk}$ originating solely from entanglement between regular matter particles.
In particular, the generation of $S_{\rm bulk}$ must involve spacetime (vacuum) degrees of freedom.
Examples of such processes include Hawking radiation and reheating after inflation.
Here we discuss another such process:\ $S_{\rm bulk}$ generated by Unruh radiation~\cite{Unruh:1976db,Unruh:1983ms} from accelerating domain walls.

\begin{figure}[t]
\centering
  \includegraphics[clip,width=0.9\columnwidth]{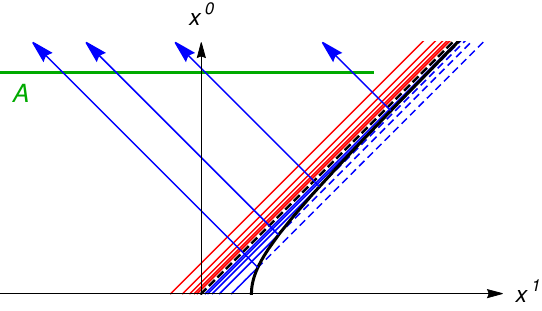}
\caption{Generation of $S_{\rm bulk}$ by an accelerating domain wall. The blue and red lines are entanglement partners of each other. This results in the region $A$, shown in green, to have a large $S_{\rm bulk}$.}
\label{fig:wall}
\end{figure}

Consider a domain wall in 4-dimensional flat spacetime which is extended in the $x^2$-$x^3$ directions and is accelerating in the $x^1$ direction.
In an inertial reference frame, the domain wall appears to emit radiation.
This occurs because the modes of a light quantum field colliding with the domain wall from behind are (partially) reflected by it, which converts these modes into semiclassical excitations on top of the vacuum; see blue arrows in Fig.~\ref{fig:wall}.
(For a review and recent analyses, see Refs.~\cite{Brout:1995rd,Akal:2020twv,Reyes:2021npy}.)

An important point is that this process stretches the wavelength of reflected modes.
In particular, radiation emitted later corresponds to a shorter wavelength mode at a fixed early time.
We postulate that, as in the case of Hawking radiation~\cite{Hawking:1974sw} and the generation of fluctuations in cosmic inflation~\cite{Mukhanov:1981xt,Hawking:1982cz,Starobinsky:1982ee,Guth:1982ec}, this picture can be extrapolated formally to an infinitely short distance, below the Planck length.
This allows for converting an arbitrary amount of short distance vacuum entanglement to entanglement involving physical radiation.
In particular, if we take a spatial region $A$ that contains the radiation but not its partner modes, then we can obtain a large contribution to $S_{\rm bulk}$ from this process.
This is illustrated in Fig.~\ref{fig:wall}.

This mechanism of generating $S_{\rm bulk}$ operates at any wall separating bubble universes.
It converts entanglement in a semiclassical vacuum, which is assumed to take the flat space form at short distances~\cite{Bunch:1978yq}, into that involving radiation emitted by the wall.
There are two classes of walls relevant for our purpose.

The first is a bubble wall separating a nucleated bubble from the ambient bubble (parent dS bubble in our context).
In this case, the bubble wall accelerates outward, so that the radiation lies inside the bubble.
This radiation is homogeneous on a Friedmann-Robertson-Walker (FRW) equal-time slice and has coarse-grained entropy density
\begin{equation}
  s \sim \left( \frac{\sqrt{-\kappa}}{2\pi a(t)} \right)^3,
\label{eq:s}
\end{equation}
where $a(t)$ is the scale factor at FRW time $t$, and $1/\sqrt{-\kappa}$ is the comoving curvature length scale at an early stage of the bubble universe, when $a(t) \approx \sqrt{-\kappa}\, t$.

The second is a domain wall separating two bubbles colliding with each other.
A domain wall relevant for our discussion is that separating the central bubble and one of the surrounding AdS bubbles colliding with it.
In this case, the domain wall accelerates outward in the AdS bubble~\cite{Freivogel:2007fx,Chang:2007eq}, so the mechanism described above applies to the AdS bubble; in Fig.~\ref{fig:wall} the regions left and right of the wall would correspond to the AdS and central bubbles, respectively.
If the domain wall is also accelerating away from the central bubble, the radiation emitted into the central bubble also results in a large $S_{\rm bulk}$, although this is not relevant for our setup.

\section{Entanglement Island from Surrounding Collapsing Bubbles}
\label{sec:main}

In this section, we argue that a sufficiently large spacelike region $R$ in the multiverse has an island $I$.
We use the method of island finder~\cite{Bousso:2021sji} to demonstrate this.
First, we locate a partial Cauchy surface $I'$ that (i) is spacelike separated from $R$, (ii) provides a reduction of generalized entropy $S_{\rm gen}(I' \cup R) < S_{\rm gen}(R)$, and (iii) has the boundary $\partial I'$ that is quantum normal or quantum antinormal with respect to variations of the generalized entropy $S_{\rm gen}(I' \cup R)$.
We will find such an $I'$ which has a quantum antinormal boundary.
We then argue that there is a partial Cauchy surface $I_0$ whose domain of dependence, $D(I_0)$, contains $I'$ and whose boundary, $\partial I_0$, is quantum normal with respect to variations of $S_{\rm gen}(I_0 \cup R)$.
Having such an $I'$ and $I_0$ guarantees the existence of a non-empty island $I$.

We focus on ($3+1$)-dimensional spacetime throughout our analysis, although it can be generalized to other dimensions.
In our analysis below, we assume that the central bubble is either a dS or Minkowski bubble, which simplifies the analysis~\cite{Freivogel:2007fx,Chang:2007eq}.
We believe that a similar conclusion holds for an AdS central bubble, but demonstrating this requires an extension of the analysis.

The argument in this section consists of several steps.
In Section~\ref{subsec:antinormal}, we identify a two-dimensional quantum antinormal surface $\partial \Sigma'$ in a surrounding AdS bubble for a region $R$ in the central bubble.
In Section~\ref{subsec:sewing}, we gather a portion of $\partial \Sigma'$ in each surrounding bubble and sew them together to form a closed quantum antinormal surface $\partial I'$ which encloses $R$.
In Section~\ref{subsec:reduction}, we argue that appending $I'$ reduces the generalized entropy of $R$ and hence it can serve as the $I'$ of Ref.~\cite{Bousso:2021sji}.
In Section~\ref{subsec:I0}, we find $I_0$, establishing the existence of a non-empty QES for $R$.
Finally, Section~\ref{subsec:inv-island} contains some discussion about the (inverted) island $I$.

While our argument applies more generally, in this section we consider a setup that involves only a central bubble and its surrounding AdS bubbles.
We discuss more general cases in Section~\ref{sec:evol}.

\subsection{Quantum Antinormal Surface in a Colliding Collapsing Bubble}
\label{subsec:antinormal}

\begin{figure}[t]
\centering
\vspace{0.2cm}
  \includegraphics[clip,width=0.9\columnwidth]{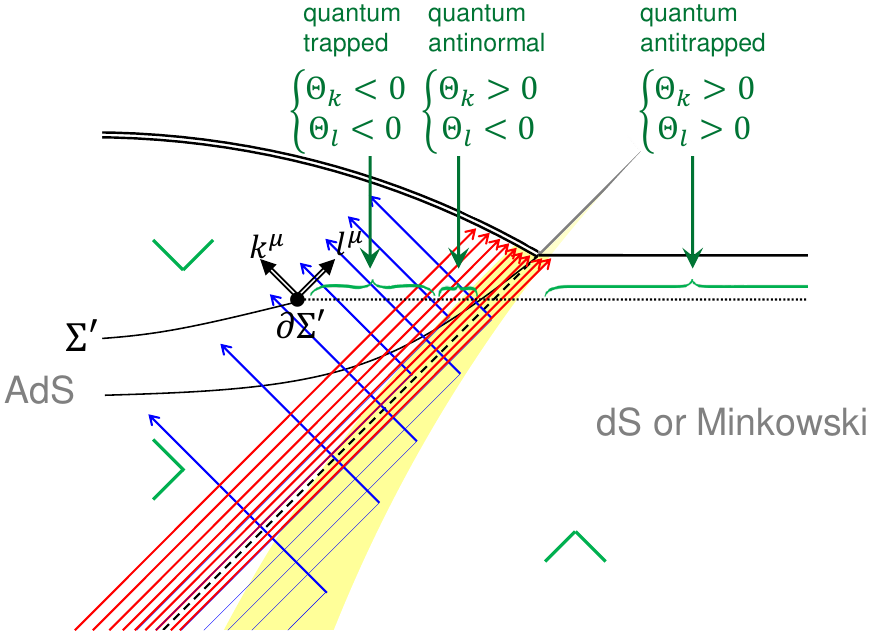}
\caption{Penrose diagram showing the region near the domain wall (yellow strip) separating the central dS/Minkowski and surrounding AdS bubbles at late times.
The transverse directions corresponding to the hyperboloid $H_2$ have been suppressed.
$\partial \Sigma'$ is a boundary of a partial Cauchy surface $\Sigma'$ and $k^\mu$, $l^\mu$ are future-directed null vectors orthogonal to it.
Blue and red arrows indicate Unruh radiation and their partner modes, respectively, and the double line at the top of the AdS bubble represents the big crunch singularity.
The signs of classical expansions $\theta_{k,l}$ are shown in green following the Bousso wedge convention~\cite{Bousso:1999xy}.}
\label{fig:Sigma}
\end{figure}

Let us consider the central bubble and only one of the surrounding AdS bubbles.
These bubbles are separated by a domain wall.
This system preserves invariance under an $SO(2,1)$ subgroup of $SO(3,1)$ symmetry of a single Coleman-De~Luccia bubble.
The spacetime is thus given by a warped product of a two-dimensional hyperboloid $H_2$ with a two-dimensional spacetime $M_2$.
Consider a two-dimensional hyperbolic surface $\partial \Sigma'$ given by the $SO(2,1)$ orbit of a spacetime point as shown in Fig.~\ref{fig:Sigma}.
We denote the partial Cauchy surface which is bounded by $\partial \Sigma'$ and extending toward the AdS side by $\Sigma'$.

We focus on the region near the domain wall at late times.
Given a $\partial \Sigma'$ in this region, let $k^\mu$ and $l^\mu$ be the future-directed null vectors orthogonal to $\partial \Sigma'$, pointing inward and outward relative to $\Sigma'$, respectively, as depicted in Fig.~\ref{fig:Sigma}.
We normalize them such that $k \cdot l = -2$ and denote the corresponding classical and quantum expansions by $\theta_{k,l}$ and $\Theta_{k,l}$, respectively.
Here, $\Theta_{k,l}$ are given by the changes in the generalized entropy $S_{\rm gen}(\Sigma' \cup R)$ under infinitesimal null variations of $\partial \Sigma'$~\cite{Bousso:2015mna}.

Suppose that a surface $\partial \Sigma'$ in the AdS bubble is located near the big crunch singularity but sufficiently far from the domain wall.
This surface is classically trapped ($\theta_k, \theta_l < 0$).
When $\partial \Sigma'$ is moved toward the central bubble, first it becomes normal ($\theta_k < 0$, $\theta_l > 0$) and then antitrapped ($\theta_k, \theta_l > 0$)~\cite{Freivogel:2007fx,Chang:2007eq}.
What about the quantum expansions?

In general, $S_{\rm bulk}$, and hence $S_{\rm gen}$, can only be defined for a closed surface, and its change $\delta S_{\rm bulk}$ under a small variation of the surface depends non-locally on the entire surface.
In our setup, however, the only relevant contribution to $\delta S_{\rm bulk}(\Sigma' \cup R)$ comes from partner modes of the Unruh radiation emitted by the domain wall into the AdS bubble, and we can locally determine the signs of $\Theta_{k,l}$.%
\footnote{The contribution from partner modes of Unruh radiation emitted into the central bubble is not relevant if $R$ is sufficiently large such that it intersects most of the radiation, since then the contribution has the same sign as the variation of the area ${\cal A}(\partial \Sigma')$.}

Suppose we locally deform $\partial \Sigma'$ in the $\pm l$ direction.
Then, $\delta S_{\rm bulk}$ receives a contribution from the reflected modes, depicted by blue arrows in Fig.~\ref{fig:Sigma}.
This contribution, however, is not strong enough to compete with the classical expansion, since the modes are spread out in the $l$ direction.

To see this explicitly, let us assume that every radiation quantum carries $O(1)$ entropy, and that the rate of emission as viewed from the domain wall's frame is controlled by the Unruh temperature $T = a_{\rm w}/2\pi$, where $a_{\rm w}$ is the acceleration of the domain wall.
We then find that%
\footnote{We thank Adam Levine for discussion on obtaining the quantum contributions.}
\begin{equation}
  \left|\delta S_{\rm bulk}\right| \sim \frac{a_{\rm w}^3 \ell^6}{\lambda r^2 (t_\infty - x^-)^3} \delta r\, \Omega_{\rm H},
\label{eq:S_bulk-1}
\end{equation}
where $\ell$ is the AdS radius in the bubble, $(t,r)$ are the location of $\partial \Sigma'$ in the coordinates~\cite{Freivogel:2007fx,Chang:2007eq}
\begin{equation}
  ds^2 = -f(r)dt^2 + \frac{dr^2}{f(r)} + r^2 dH_2^2,
\label{eq:def-r}
\end{equation}
$\delta r$ is the change of $r$ when we deform $\partial \Sigma'$ in the $l$ direction, and $\Omega_{\rm H}$ is the coordinate area of the portion of the hyperboloid for which we deform $\partial \Sigma'$.
Also, $\lambda$ is a parameter appearing in the trajectory of the domain wall
\begin{equation}
  \begin{pmatrix} t \\ r \end{pmatrix}
  \simeq \begin{pmatrix} t_\infty - t_\infty e^{-\lambda(\tau - \tau_0)} \\ r_0\, e^{\lambda(\tau - \tau_0)} \end{pmatrix},
\label{eq:wall}
\end{equation}
where $\tau$ is the proper time along the domain wall trajectory, with $r_0 = r(\tau = \tau_0)$ and $t_\infty = t(\tau = \infty)$, and we have introduced the null coordinates
\begin{equation}
  x^\pm = t \pm \frac{\ell^2}{r}.
\label{eq:null}
\end{equation}
To derive the above expressions, we have assumed that $\lambda\ell \gtrsim 1$ and $r$ is sufficiently larger than $\ell$ so that $f(r) \sim r^2/\ell^2$, which implies $t_\infty \sim \ell^2/r_0$ (also $t_\infty > \ell^2/r_0$).

The expression in Eq.~\eqref{eq:S_bulk-1} should be compared with the corresponding change in area,
\begin{equation}
  \left|\frac{\delta {\cal A}}{4 l_{\rm P}^2}\right| \sim \frac{1}{l_{\rm P}^2}\, r \delta r\, \Omega_{\rm H}.
\end{equation}
Assuming that the scalar potential responsible for the domain wall is characterized by a single energy scale $E$, we find $\ell \sim 1/E^2 l_{\rm P}$ and $\lambda \sim a_{\rm w} \sim E$,%
\footnote{The second relationship holds for generic bubbles.
For supersymmetric bubbles, we instead have $\lambda \sim a_{\rm w} \sim 1/\ell$.}
so
\begin{equation}
  \left|\frac{\delta S_{\rm bulk}}{\delta{\cal A}/4 l_{\rm P}^2}\right| \lesssim \frac{l_{\rm P}}{\ell},
\end{equation}
where we have only considered $\partial \Sigma'$ satisfying $t < t_\infty$.
We indeed find that the quantum effect, $\delta S_{\rm bulk}$, is negligible compared to the classical contribution, $\delta{\cal A}/4 l_{\rm P}^2$, for $\ell$ sufficiently larger than $l_{\rm P}$.

On the other hand, if we vary $\partial \Sigma'$ in the $\pm k$ direction, $\delta S_{\rm bulk}$ receives a contribution from the partner modes, depicted by red arrows in Fig.~\ref{fig:Sigma}.
If $\partial \Sigma'$ is far from the domain wall, this contribution is small, so that $\partial \Sigma'$ remains trapped at the quantum level:\ $\Theta_{k,l} < 0$.
However, if $\partial \Sigma'$ is moved toward the null surface to which the domain wall asymptotes, $x^+ = t_\infty$, the contribution becomes enhanced because the partner modes are squeezed there.

Specifically, the quantum effect can be estimated as
\begin{equation}
  \left|\delta S_{\rm bulk}\right| \sim \frac{a_{\rm w}^3 \ell^6}{\lambda r^2 (x^+ - t_\infty)^3} \delta r\, \Omega_{\rm H}.
\end{equation}
Here, we have assumed that the reflected modes, the partners of which $\partial \Sigma'$ crosses, all pass through $\Sigma'$, which requires
\begin{equation}
  t > t_\infty - \biggl(\frac{1-c}{1+c}\biggr) \frac{\ell^2}{r},
\label{eq:t-t_inf}
\end{equation}
where $c = (t_\infty - \ell^2/r_0)/(t_\infty + \ell^2/r_0)$ is a constant satisfying $0 < c < 1$.
We thus find that the relevant ratio is given by
\begin{equation}
  \left|\frac{\delta S_{\rm bulk}}{\delta{\cal A}/4 l_{\rm P}^2}\right| \sim \frac{\ell^5 l_{\rm P}}{r^3 (x^+ - t_\infty)^3},
\label{eq:comp-k}
\end{equation}
and the quantum effect can indeed compete with the classical contribution when $\partial \Sigma'$ approaches the null surface $x^+ = t_\infty$.%
\footnote{For supersymmetric bubbles, the numerator becomes $\ell^4 l_{\rm P}^2$. In this case, we need a more careful analysis to show that ${\delta S_{\rm bulk}}$ can compete with ${\delta{\cal A}/4 l_{\rm P}^2}$.}

Since the sign of $\delta S_{\rm bulk}$ from this effect is such that $S_{\rm bulk}$ gets reduced when $\partial \Sigma'$ is deformed in the $-k$ direction, $\Theta_k$ can become positive, making $\partial \Sigma'$ quantum antinormal:
\begin{equation}
  \Theta_k > 0, \quad \Theta_l < 0.
\label{eq:cond-1}
\end{equation}
We assume that this transition happens before $\partial \Sigma'$ changes from being classically trapped to normal.%
\footnote{If this assumption does not hold, we still have an island as will be shown in Section~\ref{subsec:I0}.}
This behavior of quantum expansions is depicted in Fig.~\ref{fig:Sigma}.

\subsection{Forming a Closed Quantum Antinormal Surface}
\label{subsec:sewing}

In the previous subsection, we have shown that there is a quantum antinormal surface $\partial \Sigma'$ in the AdS bubble.
If there were no other bubbles except for these two bubbles, then this surface would extend infinitely in $H_2$ and would have an infinite area.

\begin{figure}[t]
\centering
  \includegraphics[clip,width=0.9\columnwidth]{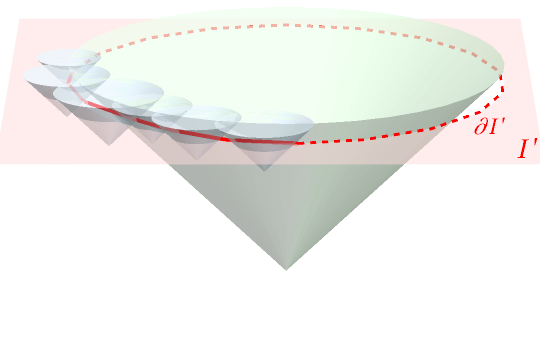}
\vspace{-1.2cm}
\caption{A sketch of the construction of closed codimension-2 surface $\partial I'$.
The central bubble and some of the surrounding AdS bubbles are depicted as the green and blue cones, respectively.
The region $I'$ is defined as a partial Cauchy surface bounded by and outside $\partial I'$.}
\label{fig:I'}
\end{figure}

However, this is not the case because the central bubble is surrounded by a multitude of AdS bubbles, as shown in Fig.~\ref{fig:I'}.
The surface $\partial \Sigma'$ corresponding to a particular AdS bubble is cut off by the domain walls resulting from collisions with the neighboring AdS bubbles.
Thus, we are left with a finite portion of $\partial \Sigma'$.
Such a finite-sized, quantum antinormal surface can be obtained in each AdS bubble, which we denote by $\sigma'_i$ ($i = 1,2,\cdots$).

These surfaces $\sigma'_i$ can be connected with appropriate smoothing in such a way that the resulting closed surface encloses the central bubble and is quantum antinormal everywhere.
To see this, we note that we have some freedom in choosing the values of $(t,r)$ for each $\sigma'_i$.
Using this freedom, we can make two adjacent $\sigma'_i$'s intersect along a curve.
The resulting ``kink'' can then be smoothed at a length scale smaller than that of bulk entanglement.
This smoothing retains quantum antinormalcy, so we end up with a closed, quantum antinormal surface.

We label this closed surface as $\partial I'$, and the partial Cauchy surface bounded by $\partial I'$ and outside it as $I'$; see Fig.~\ref{fig:I'}.
Note that $\partial I'$ being quantum antinormal means that $\Theta_k>0$ and $\Theta_l<0$, where the quantum expansions are defined using $S_{\rm bulk}(I' \cup R)$.

\subsection{Reduction of the Generalized Entropy}
\label{subsec:reduction}

We now move on to discuss the generalized entropy.
For a sufficiently large $R$, we expect that the region $I'$ reduces the generalized entropy of $R$ in the sense that%
\footnote{This implies that $I'$ violates the Bekenstein bound~\cite{Bekenstein:1980jp,Casini:2008cr}.}
\begin{equation}
  S_{\rm gen}(I' \cup R) < S_{\rm gen}(R).
\label{eq:cond-2}
\end{equation}
To understand this, we first note that Unruh radiation from the bubble walls of the central and surrounding bubbles, as well as that from the domain walls separating the central and surrounding bubbles, contributes to entanglement between $R$ and $I'$.
Appending $I'$ to $R$ therefore reduces the $S_{\rm bulk}$ contribution to $S_{\rm gen}$.

To illustrate Eq.~\eqref{eq:cond-2}, let us take $R$ to be a spherically symmetric region in the central bubble.
We assume that the distribution of AdS bubbles surrounding and colliding with the central bubble is statistically spherically symmetric.
We then append $I'$ to $R$ and compare the decrease in $S_{\rm gen}$ due to the change of $S_{\rm bulk}$ with the increase in $S_{\rm gen}$ coming from ${\cal A}(\partial I')$.

We do this comparison by focusing on an infinitesimal solid angle ${\rm d}\Omega_{\rm S}$ in the central bubble.
Using Eq.~\eqref{eq:s}, we can estimate the differential change in $S_{\rm gen}$ due to Unruh radiation from the central bubble wall to be
\begin{align}
  {\rm d}S_{\rm bulk} &\equiv \left[S_{\rm bulk}(I' \cup R) - S_{\rm bulk}(R)\right] \frac{{\rm d}\Omega_{\rm S}}{4\pi} 
\nonumber\\
  &\sim -\frac{1}{32\pi^3}\sinh\left(2\sqrt{-\kappa}\chi_*\right) {\rm d}\Omega_{\rm S},
\label{eq:del-S_bulk}
\end{align}
where $\chi_*$ is the coordinate radius of $R$ in the hyperbolic version of the FRW metric.
Here, we have used the fact that the global state is pure, so that $S_{\rm bulk}(I' \cup R) = S_{\rm bulk}(\overline{I' \cup R})$.
Moreover, we have assumed that $S_{\rm bulk}(\overline{I' \cup R})$ is sufficiently smaller than $S_{\rm bulk}(R)$ and have taken $\sqrt{-\kappa}\chi_* \gg 1$.
These conditions can be satisfied if the bubble nucleation rates in the parent bubble are small, so that the collisions with AdS bubbles occur at large FRW radii in the central bubble.

The corresponding area element of $\partial I'$ is given by
\begin{equation}
  {\rm d}{\cal A} \equiv {\cal A}(\partial I')\bigr|_{{\rm d}\Omega_{\rm H}} = r_{\sigma'_i}^2 {\rm d}\Omega_{\rm H},
\end{equation}
where $r_{\sigma'_i}$ is the location of $\sigma'_i$ in coordinate $r$ defined by Eq.~\eqref{eq:def-r}, and ${\rm d}\Omega_{\rm H}$ is the hyperbolic solid angle.
By matching the area element of the domain wall expressed in hyperbolic and FRW coordinates on the side of the central bubble, we find ${\rm d}\Omega_{\rm S} \sim {\rm d}\Omega_{\rm H}$.
This leads to
\begin{equation}
  \left|\frac{{\rm d}S_{\rm bulk}}{{\rm d}{\cal A}/4 l_{\rm P}^2}\right| \sim \frac{l_{\rm P}^2}{16\pi^3 r_{\sigma'_i}^2} e^{2\sqrt{-\kappa}\chi_*}.
\end{equation}
(To do this properly, we need to regulate the solid angle $\Omega_{\rm AdS}$ which an AdS bubble asymptotically occupies and take ${\rm d}\Omega_{\rm S}$ sufficiently small so that this area element fits within the corresponding domain wall.
We can then take the limit $\Omega_{\rm AdS}, {\rm d}\Omega_{\rm S} \rightarrow 0$ afterward.)

The radius $r_{\sigma'_i}$ is microscopic and is controlled by $l_{\rm P}$ and $\ell_i$, where $\ell_i$ is the AdS radius in the bubble in which $\sigma'_i$ resides.
When a surface $\partial \Sigma'$ is moved from an AdS bubble to the central bubble, the radius $r$ grows and becomes macroscopic.
However, this transition occurs mostly in the region where $\partial \Sigma'$ is classically normal, and since $\sigma'_i$ resides on the AdS side of it, $r_{\sigma'_i}$ is small.

We thus find that for a sufficiently large region $R$ satisfying
\begin{equation}
  \sqrt{-\kappa} \chi_* \gtrsim \log \left( \frac{4\pi^2 {\rm max}_i(r_{\sigma'_i})}{l_{\rm P}}\right),
\label{eq:cond-R}
\end{equation}
appending $I'$ to $R$ reduces $S_{\rm gen}$, so Eq.~\eqref{eq:cond-2} holds in this case.

\subsection{Existence of a Quantum Extremal Surface}
\label{subsec:I0}

The existence of a surface $\partial I'$ satisfying Eqs.~\eqref{eq:cond-1} and \eqref{eq:cond-2} is not sufficient to ensure that of a non-empty island $I$ for $R$.
The existence of an island, however, is ensured~\cite{Bousso:2021sji} if there is a partial Cauchy surface $I_0$ that (i) is spacelike separated from $R$, (ii) has the boundary $\partial I_0$ that is quantum normal with respect to $S_{\rm gen}(I_0 \cup R)$, and (iii) encloses $I'$ in the sense that $I' \subset D(I_0)$.

\begin{figure}[t]
\centering
  \includegraphics[clip,width=1\columnwidth]{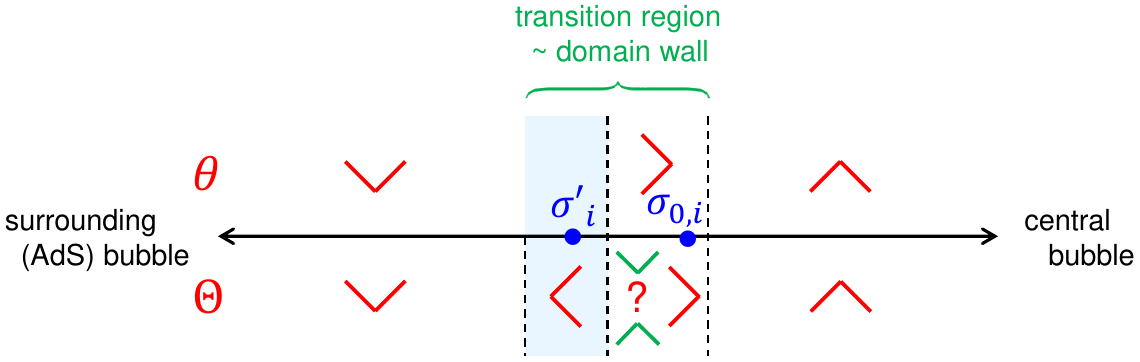}
\caption{Variations of classical and quantum expansions, $\theta$ and $\Theta$, when a two-dimensional surface $\partial \Sigma$ extending in the direction of hyperboloid $H_2$ is moved between the surrounding AdS and central bubbles.
A finite area surface $\sigma'_i$ ($\subset \partial \Sigma'$), which constitutes a portion of $\partial I'$, is taken in the regime where the surface is quantum antinormal.
A surface $\sigma_{0,i}$ ($\subset \partial \Sigma_0$) which gives $\partial I_0$ after smoothing is in the quantum normal region.}
\label{fig:theta}
\end{figure}

To argue for the existence of such $I_0$, let us consider a codimension-2 surface $\partial \Sigma_0$ similar to $\partial \Sigma'$.
Such a surface is specified by the coordinates $(t,r)$ in Eq.~\eqref{eq:def-r}.
The analysis in Sections~\ref{subsec:antinormal} and \ref{subsec:sewing} then tells us that when $\partial \Sigma_0$ is moved from the near singularity region to the central bubble, it changes from being quantum trapped to quantum antinormal (as viewed from the side opposite to the central bubble, which we denote by $\Sigma_0$).
This occurs before the classical expansions become normal.
As we move the surface further, we expect that the quantum effect becomes subdominant at some point, making the signs of quantum expansions the same as those of classical expansions.
In Fig.~\ref{fig:theta}, we depict possible behaviors of quantum expansions in this region by green Bousso wedges which are consistent with the quantum focusing conjecture~\cite{Bousso:2015mna}.
We can thus take $\partial \Sigma_0$ in the quantum normal region to construct the surface $\partial I_0$.

Like $\partial \Sigma'$, the surface $\partial \Sigma_0$ is truncated by AdS-AdS domain walls and becomes a finite surface $\sigma_0$.
As earlier, we form a closed surface using these truncated surfaces $\sigma_{0,i}$ ($i = 1,2,\cdots$) from each surrounding AdS bubble.
By using the freedom of locating each surface, these pieces can be sewn together to form a closed surface enclosing the central bubble.

The resulting surface, however, has folds at the junctions between AdS bubbles, with angles opposite to those required for quantum normalcy.
Nevertheless, the effect of these angles is suppressed by $O(\ell_i/r)$ compared to that of the expansions of $\sigma_{0,i}$'s in the interior of the AdS bubbles.
Therefore, by locating $\sigma_{0,i}$'s at large $r$, we can smooth out the folds to form a closed surface that is classically normal and hence quantum normal.

This surface can play the role of $\partial I_0$:
\begin{equation}
  \bigcup_i \sigma_{0,i} \xrightarrow[\mbox{\tiny smoothing}]{} \partial I_0,
\end{equation}
where we define $I_0$ as a partial Cauchy surface bounded by and outside $\partial I_0$.
It is easy to see that the smoothing can be done such that the resulting $I_0$ is spacelike separated from $R$ and $I' \subset D(I_0)$.
This guarantees the existence of an island for $R$.

We note that the existence of $I_0$ is sufficient by itself to ensure the existence of an island if $R$ is very large, satisfying Eq.~\eqref{eq:cond-R} with ${\rm max}_i(r_{\sigma'_i})$ replaced with the radius of $I_0$.
Our argument involving $I'$, however, indicates that the island exists for much smaller $R$.

\subsection{Inverted Island and Entanglement Castle}
\label{subsec:inv-island}

Given that the collisions between the central and surrounding bubbles play an essential role in the existence of $I'$ and $I_0$, we expect that $\partial I$ is located in the region near the corresponding domain walls.
In fact, it is reasonable to expect that the two possibilities for quantum expansions depicted in Fig.~\ref{fig:theta} are both realized, depending on the path along which a codimension-2 surface $\partial \Sigma$ is moved.
The edge of island $\partial I$ would then lie at the point where trajectories of $\partial \Sigma$ bifurcate to behave in these two different ways.
The structure of the Bousso wedges around this location is indeed consistent with $\partial I$ being a quantum maximin surface~\cite{Wall:2012uf,Akers:2019lzs}.

Strictly speaking, this only implies that the surface $\partial I$ is a QES.
In order for this surface to be the boundary of an island, it must be the minimal QES.
We assume that this is the case, which is true if $R$ has only one nontrivial QES with $S_{\rm gen}(I \cup R) < S_{\rm gen}(R)$.

Since the topology of $I$ is the same as that of $I'$ or $I_0$, the island $I$ for region $R$ is an {\it inverted island}, and hence does not geographically look like an island.
Let $\Xi$ be a Cauchy surface containing $R$ and $I_\Xi = D(I) \cap \Xi$ the section of the inverted island on this surface.
Given the geography, we may refer to the region $\overline{I_\Xi}$, complement of $I_\Xi$ on $\Xi$, as an {\it entanglement lake}.
However, $R$ occupies a significant portion of $\overline{I_\Xi}$, so (regarding $R$ as a land as other authors do) the region $\overline{R \cup I_\Xi}$ which corresponds to water is more like a moat; see Fig.~\ref{fig:castle}.
In this sense, the region $\overline{I_\Xi}$ in the present context may be called an {\it entanglement castle}.

\section{Cosmological Evolution}
\label{sec:evol}

Consider a Cauchy surface $\Xi$ in the global spacetime.
The existence of a non-empty island $I$ for a subregion $R$ of $\Xi$ implies that the information about the semiclassical state in $I_\Xi = D(I) \cap \Xi$ is encoded in the fundamental degrees of freedom associated with $R$.
Therefore, physics at the semiclassical level can be fully described by the fundamental degrees of freedom associated with the partial Cauchy surface $\overline{I_\Xi} = \Xi \setminus I_\Xi$.

In the eternally inflating multiverse, an inverted island $I$ appears for sufficiently large $R$.
This implies that the semiclassical physics of the multiverse, which is all that we need to make cosmological predictions, is described by the fundamental degrees of freedom in a finite volume portion of a Cauchy slice that involves $R$.
We call such a surface an {\it effective Cauchy surface}.

Here we make two general comments about effective Cauchy surfaces.
First, the location of the island $D(I)$, or $\partial I$, depends on the Cauchy surface.
For example, since $R$ is spacelike separated from $I$, a Cauchy surface describing the state of the parent bubble cannot have $\partial I$ around the central bubble as seen in the previous section.
However, in this case there exists a region $R_{\rm p}$ in the parent bubble such that an island $I_{\rm p}$ appears around the parent bubble, so that the effective Cauchy surface is given by $\Xi \setminus (D(I_{\rm p}) \cap \Xi)$.
In general, when we consider a Cauchy surface describing the state of an earlier bubble, the relevant island appears around that bubble.

Second, when two or more (non-surrounding) bubbles collide, we may want to consider Cauchy surfaces spanning all of these bubbles to describe the collision.
In this case, we can choose a region $R_{\rm c}$ spanning the colliding bubbles such that the island $I_{\rm c}$ encloses all the colliding bubbles.
This allows us to describe the bubble collision directly without relying on reconstruction from microscopic information in the fundamental degrees of freedom in $R$.

\begin{figure}[t]
\centering
\vspace{0.1cm}
  \includegraphics[clip,width=0.9\columnwidth]{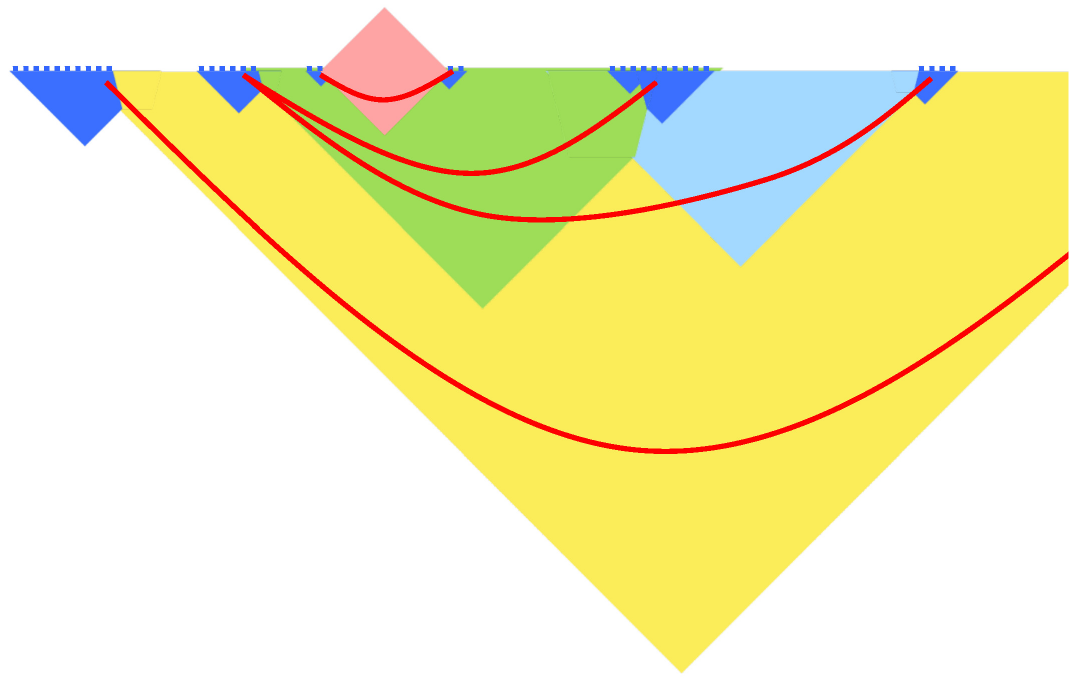}
\caption{Several effective Cauchy surfaces for a given geometry are depicted by red lines.
A microstate of the fundamental degrees of freedom on an effective Cauchy surface can describe the full semiclassical physics of the multiverse.}
\label{fig:slices}
\end{figure}
A sketch of the global multiverse illustrating the above points is given in Fig.~\ref{fig:slices}, where possible effective Cauchy surfaces are depicted by red lines.
For a given gauge choice, the state on an effective Cauchy surface $\Upsilon_1$ can uniquely determine the state on an effective Cauchy surface $\Upsilon_2$ that is in the future domain of dependence of $\Upsilon_1$.
In general, the final state of this time evolution is given by a superposition of states in different geometries ${\cal M}_i$:
\begin{equation}
  \ket{\Psi(\Upsilon_1)} \xrightarrow[\mbox{\tiny evolution}]{\mbox{\tiny time}} \sum_{i \in \mbox{\tiny geometries}}\!\!\! c_i \ket{\Psi(\Upsilon_{2,i})}_{{\cal M}_i}.
\label{eq:evol}
\end{equation}
Here, all ${\cal M}_i$'s share the surface $\Upsilon_1$ and the state on it, and $\Upsilon_{2,i}$ is an effective Cauchy surface on the geometry ${\cal M}_i$ which is in the future domain of dependence of $\Upsilon_1$.

It is worth noting that the evolution equation in Eq.~\eqref{eq:evol} takes the form that once the knowledge of the current state, $\ket{\Psi(\Upsilon_1)}$, is given, we can predict its future, more precisely what an observer who is a part of the state can in principle see in their future.
Note that the equation does not allow us to infer from $\ket{\Psi(\Upsilon_1)}$ the global state of the multiverse in the past.
This structure is the same as time evolution of states in the Schr\"{o}dinger picture of quantum mechanics.

Our approach solves the measure problem in the sense described above:\ once we are given the initial state on an effective Cauchy surface, we can in principle predict any future observations.
The existence of the inverted island implies that the necessary information for this prediction, i.e.\ the physics of matter excitations over semiclassical spacetimes, is fully encoded in the microstate of the fundamental degrees of freedom associated with the effective Cauchy surface.
As discussed in Ref.~\cite{Nomura:2019qps} for a dS spacetime, this information is expected to be encoded in quantum correlations between the matter and Unruh radiation degrees of freedom.

\section{Conclusions}
\label{sec:concl}

In this paper, we have shown that a Cauchy surface $\Xi$ in an eternally inflating multiverse has an entanglement island for a sufficiently large subregion $R \subset \Xi$.
The island $I_\Xi$ on $\Xi$ is, in fact, an inverted island surrounding the region $R$, implying that the semiclassical physics of the multiverse is fully described by the fundamental degrees of freedom associated with the finite region $\overline{I_\Xi}$, the complement of $I_\Xi$ on $\Xi$.
This provides a regularization of infinities which caused the cosmological measure problem.

As in the case of a black hole, the emergence of an island is related to the existence of a singularity in the global spacetime; in the multiverse, this role is played by the big crunch singularities in the collapsing AdS bubbles.
This picture is consistent with the interpretation of singularities in Refs.~\cite{Nomura:2018kia,Nomura:2019qps,Nomura:2020ska}:\ their existence signals that a portion of the global spacetime is intrinsically semiclassical, arising only as an effective description of more fundamental degrees of freedom associated with other spacetime regions.

The result in this paper strongly suggests the existence of a description of the multiverse on finite spatial regions.
Proposals for such descriptions include Refs.~\cite{Freivogel:2006xu,Susskind:2007pv,Sekino:2009kv} and Refs.~\cite{Nomura:2011dt,Nomura:2011rb,Nomura:2016ikr} in which the fundamental degrees of freedom are associated with the spatial infinity of an asymptotic Minkowski bubble and the (stretched) cosmological horizon, respectively.
It would be interesting to explore precise relations between these holographic descriptions and the description based on the global spacetime presented in this paper.

\acknowledgments

We thank Raphael Bousso, Adam Levine, and Arvin Shahbazi-Moghaddam for useful conversations.
This work was supported in part by the Department of Energy, Office of Science, Office of High Energy Physics under contract DE-AC02-05CH11231 and award DE-SC0019380 and in part by MEXT KAKENHI grant number JP20H05850, JP20H05860.

\end{document}